\documentclass[%
 reprint,
groupedaddress,
 amsmath,amssymb,
 aps,
]{revtex4-2}

\usepackage{graphicx}
\usepackage{dcolumn}
\usepackage{bm}

\begin{document}

\preprint{APS/123-QED}


\title{Spatially correlated rotational dynamics reveals strain dependence in amorphous particle packings}

\author{Dong Wang}
\thanks{These authors contributed equally}
\affiliation{Department of Physics \& Center for Non-linear and Complex Systems, Duke University, Durham, North Carolina 27708, USA}
\affiliation{Department of Mechanical Engineering \& Materials Science, Yale University, New Haven, Connecticut 06520, USA}

\author{Nima Nejadsadeghi}
\thanks{These authors contributed equally}
\affiliation{Mechanical Engineering, University of Kansas, 1530 W. 15th St. Lawrence, KS 66045-7609, USA}

\author{Yan Li}
\thanks{These authors contributed equally}
\affiliation{Computer Science \& Engineering, University of Minnesota, Minneapolis, MS, USA}

\author{Shashi Shekhar}
\email{shekhar@umn.edu}
\affiliation{Computer Science \& Engineering, University of Minnesota, Minneapolis, MS, USA}

\author{Anil Misra}
\email{amisra@ku.edu}
\affiliation{Civil, Environmental \& Architectural Engineering, University of Kansas, 1530 W. 15th St. Lawrence, KS 66045-7609, USA}

\author{Joshua A. Dijksman}
\email{joshua.dijksman@wur.nl}
\affiliation{Physical Chemistry and Soft Matter, Wageningen University \& Research, Stippeneng 4, 6708 WE Wageningen, The Netherlands}

\date{\today}

\begin{abstract}
Microstructural dynamics in amorphous particle packings is commonly probed by quantifying particle displacements. While rigidity in particle packings emerges when displacement of particles are hindered, it is not obvious how the typically disordered displacement metrics connect to mechanical response. Particle rotations, in contrast, are much less sensitive to confinement effects, while still sensitive to the mechanics of the packing. So far, little attention has been paid to connect microscopic rotational motion to mechanics of athermal amorphous packings. We demonstrate through experimental measurements that particle packing mechanics can be directly linked to the rotational motion of even round particles in a sheared packing. Our results show that the diffusive nature of rotational dynamics is highly strain sensitive. Additionally, there is substantial spatial correlation in rotation dynamics that is a function of the particle friction and packing density. Analysis of our measurements reveals that particle rotation dynamics plays an essential role in amorphous material mechanics.
\end{abstract}

\maketitle


\section{\label{sec:level1}Introduction}
Amorphous packings of particles occur in many contexts, ranging from glassy polymers to colloidal gels and geological sediments. These materials are well known to have complex deformation behavior. For example, their mechanical response is often strain history dependent~\cite{1977_lade,2002_viasnoff,2009_lee,tighe2014shear}. The amorphous nature of the microstructure of these systems makes it notoriously challenging to understand the origin of such strain history dependence and mechanical behavior in general~\cite{berthier2011dynamical}. Notably, these ``granular'' material systems are characterized by a length-scale proportional to particle size, that makes their theoretical description using classical continuum physics concepts particularly challenging. For accurate descriptions of amorphous packing mechanics, the traditional views of continuum mechanics desperately needs updating from a microscopic point of view. One route towards a more general continuum description considers material point rotations and has an origin in the work from the Cosserat brothers~\cite{cosserat1909theory}. Indeed, it has long been recognized that the rotational motion of particles in thermally driven amorphous packings can be linked to slowdown effects and glassy dynamics~\cite{1984_schwartz, 1994_Stillinger, 2012_Edmond}. Nevertheless, much progress is needed to include (particle) rotational degrees-of-freedom in continuum mechanics approaches ~\cite{2003_matsushima, poorsolhjouy19jmps}. Here we show experimentally that even in an completely athermal amorphous packing, rotational degrees-of-freedom are directly coupled to its mechanical response, both at the particle level and via mesoscopic spatial anticorrelations in the rotation field. Our data suggests that particle rotations are an essential yet overlooked kinematical quantity in the study of dense amorphous packing.  In addition, the spatial autocorrelation analysis of rotations can reveal essential features in materials science of a large variety of materials with intrinsic length scales.

To study the role of rotational degrees of freedom, decoupling rotation from translation is challenging. In many circumstances the involved molecules, colloids or grains are not spherically symmetric, hence their rotation requires also \emph{spatial displacement} of their neighbors, particularly for high density, jammed granular materials. To probe the role of \emph{only} the rotational degrees of freedom in the strain dependence of amorphous packings, athermal round particle systems are an optimal prototypical choice. Such particles can be designed to experience contact friction, which directly couples rotational degrees of freedom to displacements. In an athermal packing of frictional disks, shear for even circular particles is thus directly coupled to rotations without necessarily requiring particle displacements.  While rotational dynamics of spherical particles in athermal packings has been probed via wave-propagation measurements~\cite{2011_merkel}, particle-level experimental evidence that links rotational degrees of freedom directly to mechanics in amorphous packings has so far not been obtained. The unique set of experimentally measured data analyzed in this work shows that the particles' micro-rotation dynamics are linked to both the packing density and the particle surface characteristics (or ``friction'') which directly mediates in tangential motion of contacting circular grain-pairs. We will see that particle rotations display strain-induced diffusive behavior even at very small strain amplitudes. The diffusive behavior changes with particle packing density and friction coefficient in manners consistent with previously observed packing properties~\cite{ren13prl, 2018_trimer_dong, 2020_wang}. Additionally, particle rotations display non-local correlations as revealed by spatial auto-correlation measures~\cite{moran1950notes}, indicating that the non-local mechanical effects well known to exist in sheared glassy granular media and amorphous materials in general~\cite{hebraud1998modecoupling,2009_bocquet, 2013_henann} can be mediated by rotational dynamics.  Our results show that rotational degrees of freedom are a crucial element to be considered in the quest to understand the flow behavior of amorphous materials.

Granular material systems exhibit many non-standard physical phenomena, such as that of negative group velocity \cite{nejadsadeghi2020role,WEI2020105433}, frequency band gaps \cite{MOURAILLE2008498,boechler2011tunable,goncu2012exploiting,misra16cmt}, chirality~\cite{misra2020chiral} and load path dependency. The latter is the key physical phenomenon that underlies our results: collections of discrete athermal particles have the remarkable property that they can  become \emph{rigid} when assembled into certain arrangements. Referred to as granular media, these loose particle packings can resist shear or compression when a sufficient number of them is present per unit volume, a feature commonly called \emph{jamming}. Even when not spatially confined, packing of particles can enhance its rigidity when the assembly is subject to shear strain. This shear deformation induced rigidification is known as \emph{shear jamming}. The peculiar strain dependence of granular media can be indicated by the correspondence between rotational particle motion and collective packing mechanics. Indeed, some work had already hinted at the relevance of rotational degrees of freedom of particles in loose particle assemblies for both slow~\cite{misra1997measured,2003_matsushima, kuhn2004contact, ando2012grain} and fast granular flows~\cite{seto13prl,lin2015hydrodynamic, abhi_2020_roll_friction} and recently also for the statistical mechanics of sheared packings~\cite{2012_ciamarra,sun2020frictioncontrolled}.

\section{Rotation Kinematics}

\begin{figure}
\centering
\includegraphics[width=\linewidth]{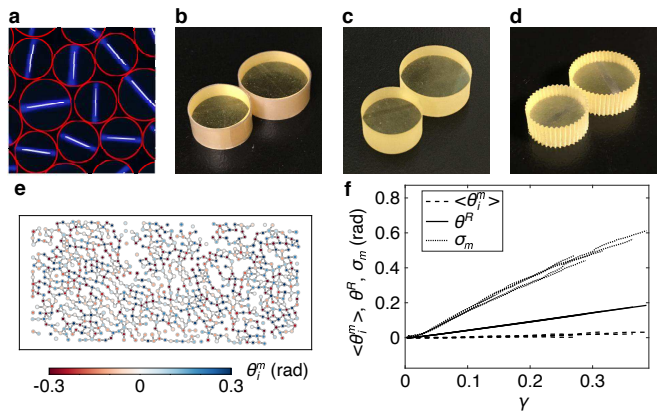}
\caption{(a) Examples of the UV image taken to track particle orientation. Blue bars show actual UV marks and white lines indicate tracked orientation. Red circles mark edges of disks. (b - d) Examples of photoelastic particles used for different inter-particle friction coefficients $\mu_{l,m,h}$ respectively. (e) The microrotation of grains induced by the first shear step; particle locations shown in their initial configuration. Results obtained for $\gamma = 0.27$, $\mu_m$ with packing fraction $\phi$=0.816. (f) evolution of the mean $\langle\theta_i^m\rangle$ and standard deviation $\sigma_m$ of the microrotation, and the rigid body rotation $\theta^R$, for five shear tests at a given density. $\theta^R$ grows linearly at a rate of 0.0013~rad per frame as expected from the imposed strain.}
\label{fig:example}
\end{figure}


Packing of disk-shaped particles at different packing fractions $\phi$, were subjected to quasi-static stepwise shear to observe the diffusive rotational dynamics of particles. The packing were imaged after each strain step. For details of the experiments, see the Methods section. In our experiments we have used quasi-two dimensional packing, as three dimensional shear experiments have the propensity for formation of finite shear localization bands into which the particle rotations typically concentrate ~\cite{cheng2019quantification,hall2010discrete,alshibli2006microscopic,ando2012grain}. In contrast, our two dimensional geometry with articulated base allows us to suppress shear localization entirely~\cite{ren13prl,2018_trimer_dong}. A fluorescent bar placed on the particles allows us to track the absolute orientation of every particle in the packings; see Fig.~\ref{fig:example}a. We use three different particle friction coefficients and refer to these as $\mu_l$, $\mu_m$ and $\mu_h$; examples of these particle types are shown in Fig.~\ref{fig:example}b-d. We probe the dynamics of orientations obtained from image analysis of each frame.  An example rotation field from a single shear step is shown in Fig.~\ref{fig:example}e for $\mu_m$ particles at packing fraction $\phi=0.816$. While the grain displacements tend to follow the imposed macro-scale deformation field, there exists fluctuations from the imposed linear macro-scale deformation field in the grain centroid displacements that follow a complex process dictated by not only the nearest or contacting grains, but also by their neighbors and by extension an increasingly larger neighborhood for every grain~\cite{MISRA20171,nima2020MMS, 2020_wang}.


\begin{figure}
\includegraphics[width=\linewidth]{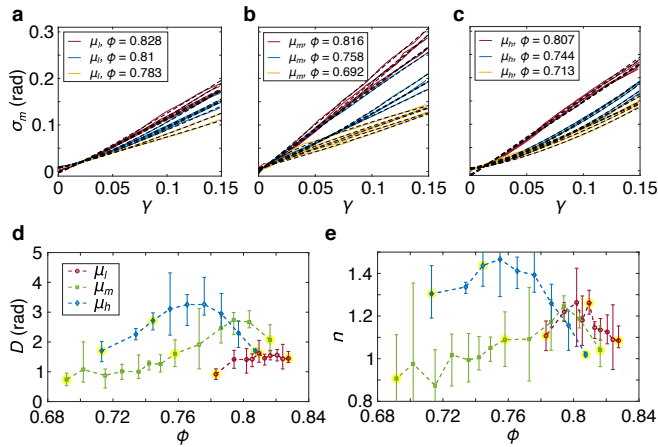}
\caption{\label{fig:Fig_fit15} Standard deviation of rotations as a function of imposed shear strain in three different packing fractions for the (a) $\mu_l$, (b) $\mu_m$, and (c) $\mu_h$ particles. All experiments are repeated five times. (d) and (e) show, respectively, the variation of the parameters $D$ and $n$ as a function of packing fraction of the sets in $\mu_{l,m,h}$. The highlighted data in yellow are the data used in panels a-c and Fig.~\ref{fig:var-global}.}
\end{figure}

Similarly, the grain rotation observed in Fig.~\ref{fig:example}e can be decomposed into two parts. One part of the rotation of each grain is a result of the imposed affine macro-scale deformation field, which contributes an overall rigid body rotation. The second rotation contribution is due to the micro-scale phenomena such as individual grain spin that we call \emph{microrotation}. Denoting the rotation of grain $i$  by $\theta_i$ and the macro-scale rigid body rotation by $\theta^R$, the microrotation of grain $i$, $\theta_i$, is obtained as $\theta_i^m = \theta_i-\theta^R$. The mean of the microrotations $\langle\theta_i^m\rangle$ and the standard deviation of microrotations $\langle{\theta_i^m}^2\rangle = \sigma_{m}$ change as a function of strain as observed in Fig.~\ref{fig:example}f for all five repeats done for $\mu_m$ at $\phi=0.816$. The initial frame is taken as the reference configuration to obtain the evolution of the grain-spin measures as a function of imposed strain. The rigid body rotation $\theta^R$ grows linearly with strain as expected from the linearly increasing strain field imposed on the packing. Notably, the mean microrotation $\langle\theta_i^m\rangle$ is zero for the packing: that is, there is no preferred direction in which grain rotation \emph{fluctuations} occur. This null-result is highly reproducible between different repeats of the experiments and consistent with earlier numerical simulations~\cite{aharonov2002shear, kuhn2004contact,calvetti1997experimental,misra1997measured,alshibli2006microscopic}. It is also noteworthy that the microrotations follow a nearly Gaussian distribution for all cases (see results in SI). However, the amplitude of grain rotation fluctuations increases strongly with strain. Note that some of such shear induced rotational fluctuations have been observed in the experimental work of Matsushima in~\cite{2003_matsushima} on nonspherical particles.

Focusing further on the growth of $\sigma_m(\gamma)$ we see that its strong growth with $\gamma$ and the reproducibility among different initial configurations is also observed for different $\phi$ over the entire range of relevant densities and $\mu$ as shown in Fig.~\ref{fig:Fig_fit15}a-c. Up to a strain of 0.15, $\sigma_m$ can be well described by the empirical relation $\sigma_m = D\gamma^n+\sigma_0$ as shown by the good quality of the fits. It is tempting to interpret prefactor $D$ as a diffusion constant as done previously for rotations induced by thermal fluctuations~\cite{kim2011colloidal, shanda2017decoupling}. Power law index $n$ indicates the (weakly) nonlinear strain dependence and $\sigma_0$ is a possible offset that is negligible for all experiments. We see that the strength of the fluctuations captured by $D(\phi)$ and $n(\phi)$ is very sensitive to friction $\mu$ as shown in Fig.~\ref{fig:Fig_fit15}d,e. The friction dependence does however capture the mechanical performance of the packing as well: at large $\mu$, particle interactions associated with rotation are stronger even at smaller $\phi$, and this trend is observed in both $D(\phi)$ and $n(\phi)$. When considering $D$ as a diffusion constant, its thermal analogue would be given by the ratio of thermal fluctuations and viscous damping. Such competition can also be seen in the rotational diffusion: both $D$ and $n$ indicate that there are two mechanisms that play a role in the rotational diffusion, which is especially visible for $\mu_m$. Initially, $D,n(\phi)$ grows with $\phi$, indicating the enhanced particle interactions that give more fluctuations in rotations. However, above a certain $\phi_c(\mu)$, $D$ decreases, and above the packing fraction, $\phi \approx 0.80$, parameters $D,n$ tend towards a plateau, indicating that competing mechanism emerge in high packing fractions suppressing the growth of further rotational fluctuations. Steric hindrance does play a role for $\mu_h$, the gear-shaped particles, but less so for the much smoother $\mu_{l,m}$ particles.  The exponent for the $\mu_h$ can be as high as  1.4, indicating superdiffusive behavior if we consider $\gamma$ as a time variable and $\langle{\theta_i^m}^2\rangle = \sigma_{m}$ as a displacement fluctuation metric. Interestingly, the values for $n(\phi)$ become \emph{independent} of $\mu$ above a volume fraction of about 0.81, tending towards a linear behavior at very high packing fraction. These trends are even more visible if we consider less strain, see Supplementary Information.

\begin{figure}
    \centering
    \includegraphics[width=\linewidth]{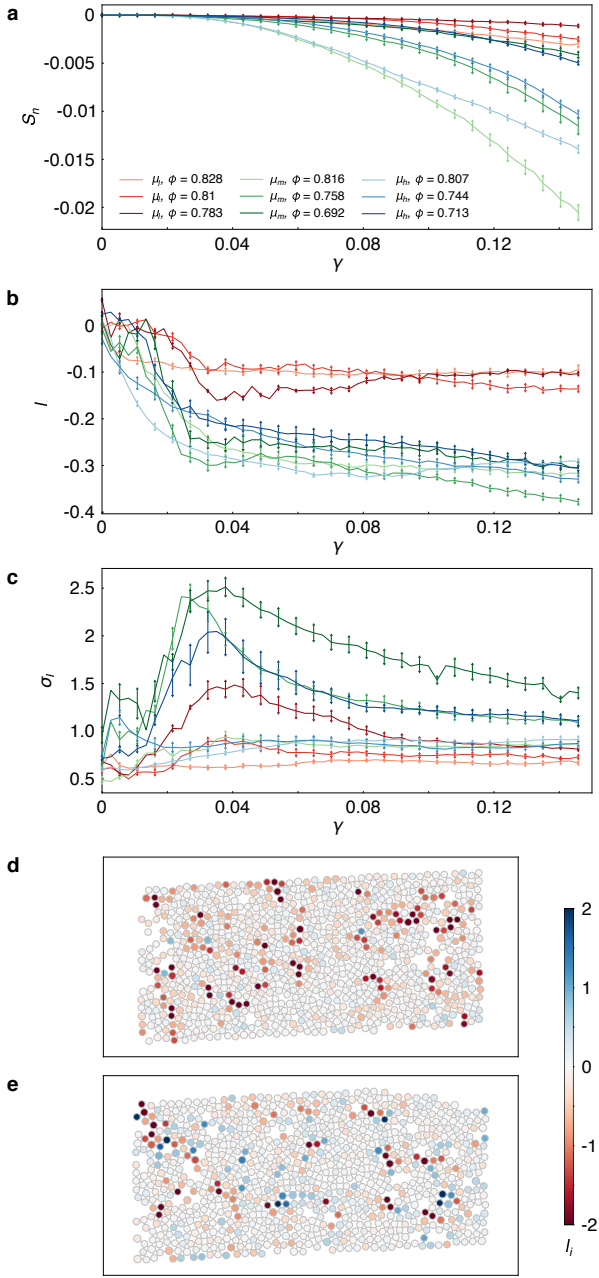}
    \caption{(a) The average neighborhood variance $\sigma_m(\gamma)$; different colors indicate different $\mu$; different hues represent different $\phi$; color scheme applies to all panels. (b) $I$ as a function of $\gamma$. (c) $I_i$ standard deviation $\sigma_I$ as as function of $\gamma$. (d) $\mu_l$, $\phi = 0.828$ local Moran's I. (e) $\mu_h$, $\phi = 0.807$ local Moran's I.}
    \label{fig:var-global}
\end{figure}

\section{Correlations in micro-rotations}
The observed Gaussian nature of the particle rotation fluctuations belies the underlying correlations in particle motion as expected to exist in dense amorphous packings where many grains are in contact. There are long range correlations in particle rotations as visible in our two step approach to quantifying the spatial autocorrelation of rotations, leading to a mathematically well defined quantitative system average signal widely used for geographical data called Moran's $I$ ~\cite{moran1950notes}. We first compute the particle average neighborhood rotational variance $S_{n}$. For details, see the Methods sections. A positive/negative value of $S_{n}$ means that in general a particle rotates in the same/opposite direction as its neighborhood. Fig.~\ref{fig:var-global}(a) shows the average neighborhood variance of materials with different $\mu$ and packing fraction $\phi$. The average neighborhood variance is clearly monotonically dependent on both $\mu$ and $\phi$ and grows with $\gamma$: the higher friction or density, the lower the average neighborhood variance, which means greater difference between a particle's micro-rotation and its neighborhood's. This anticorrelation makes mechanical sense: gear-like motion forces rotation of opposite direction in interlocking particles.

A large absolute value of $S_{n}$ may however be caused by either (dis)similarity between neighborhood particle micro-rotations, or a large variance in the micro-rotation. To focus only on the comparison of the dissimilarity among neighborhood particles micro-rotation across the packing, we have to normalize $S_{n}$ by the variance of the particle micro-rotation, $\sigma_m^2$. We showed the dynamics of $\sigma_m^2$ and its non-monotonic dependence on friction and packing fraction in Fig.~\ref{fig:Fig_fit15}. By computing $S_{n}/\sigma_m^2$, we arrive immediately at the system wide spatial autocorrelation metric called Moran's $I$. Generally, the micro-rotations of the grains in all the materials in these analyses are negatively autocorrelated: the grains rotate like a chain of gears to some extent. Figure \ref{fig:var-global}(b) shows the trends of $I$ as the shear strain increases. The differences in behavior for $\mu_{l,m,h}$ is evident: low friction particles have a weak spatial autocorrelation, whereas particles with higher friction coefficient develop stronger autocorrelations, with $I$ decreasing to -0.3. The difference between the packings with a different $\phi$ is small but not insignificant. In general, anti-autocorrelations increase with larger packing fractions. Strikingly, also the rotational correlations are very strain sensitive, with 3\% strain being enough to indicate significant difference between packings of the different $\phi$ and $\mu$.

We go one step further and use the normalized neighborhood variance to gain insight into the local mechanics of sheared amorphous packings. Spatial autocorrelations as captured by $I$ are not the same everywhere; in fact there are clusters of (anti)correlated rotations in $I_i$. We can quantify the local variability of these correlations by computing the standard deviation $\sigma_I$ of $I_i$. This metric captures the rotational ``floppyness'' in the packing: at large $\phi$ in a highly overconstrained system, interlocking grains must all have the same rotational behavior so the variability of $I$ in the packing should be small. At smaller $\phi$, there are more ways to reach mechanical equilibrium, hence the variability among correlations should be higher. Similarly, $\sigma_I$ should express the phenomena of shear jamming: at small strain, the shear jamming mechanisms has not been activated yet, so $\sigma_I$ is small. As strain increases, the packing moves from partially to completely constrained and should thus achieve a small $\sigma_I$. Finally, the role of $\mu$ should also be non-linear: at small and large $\mu$, the rotational variability should be high as per previous arguments, so $\sigma_I(\mu)$ should have an optimum. We observe indeed all these mechanically reasonable trends in $\sigma_I(\phi,\gamma,\mu)$. $I_i$ standard deviation is strain dependent, exhibiting a distinct peak floppiness at about 3\% strain. Note that at these strain levels, system level pressure is undetectable, highlighting again the sensitive nature of rotations. The connection to the rotational diffusion is also still visible in the fluctuations of the anticorrelated micro-rotation: observe how for $\phi > 0.80$, $\sigma_I$ is small for all $\mu$, precisely where also the diffusivity of rotation becomes independent of $\mu$. Finally, we show two examples of the spatial distribution $I_i(x,y)$ for two situations $\mu_l$, $\phi = 0.828$ and $\mu_h$, $\phi = 0.807$ in Fig.~\ref{fig:var-global}d,e. These examples clearly show clusters of isotropic and and anisotropic shapes emerging along boundaries and in the bulk of the packing. Spatial fluctuations can span up to ten particle diameters and can be string-like or globular, highlighting again the spatial anisotropy that can build up in the amorphous system (see Supplementary Information videos). While the complete spatial dynamics is neighborhood rotation similarity is challenging to interpret due to the dual and non-monotonic role of both friction and density, we can clearly evidence particle rotation becoming an essential parameter necessary to include in continuum modeling theories with non-local mechanical couplings inside sheared amorphous packings.  

\section{Conclusions}
We have shown that simple shear induces spatially correlated fluctuations in the a rotational dynamics of round, frictional particles. Individual particle motion is diffusive, and diffusive motion is $\mu$ and $\phi$ dependent as one would expect based on the mechanical characteristics of the packing. The local neighborhood of particles shows on-average  anticorrelated motion that reveals that two distinct mechanisms affect the mechanics of individual grains. Rotational motion fluctuations indicate the state of the system early in the deformation regime after a few percent shear strain, even though the \emph{average} particle micro-rotation is zero. Our results indicate that rotational motion is a highly relevant field in the study of amorphous particulate materials, ranging from sands to frictional emulsions, colloids and even molecular glasses. Beyond materials analyses, the results have a broader relevance to spatial data science, particularly in reference to the ``first law of Geography''~\cite{tobler1970computer} stating that nearby things are similar. The value of the widely used geographical spatial autocorrelation measure Moran's $I$ is negative for granular materials systems with a clear physical interpretation related to particle friction. This finding is in contrast with a large majority of spatial datasets coming from human-scale natural systems which have positive spatial autocorrelation. 
Intriguingly, the role of absolute interparticle orientations has long been recognized for system mechanics: the role of the bond angle is recognized as essential in constraint counting approaches for glassy polymeric systems~\cite{1995_thorpe} and is also relevant for protein folding dynamics~\cite{Jacobs2002}. Not surprisingly rotational dynamics has been measured indirectly on a system scale via dielectric spectroscopy~\cite{madden1984consistent}, for example to probe glassy dynamics in rotational degrees of freedom in nonspherically symmetric glassforming molecules. Note that not only friction can make the rotational degree of freedom relevant for the packing dynamics. Rotations also play a role for particle packings that are composed of aspherical, adhesive or deformable particles, which covers many types of particulate materials, ranging from granular materials to colloids, proteins~\cite{haradisan2019rotationalprotein}, emulsions and even metamaterials in which the node hinges are not ideal~\cite{misra2020chiral}. In particular, it is of interest to explore how energy is stored in sheared granular packings and how rotations and friction in contacts play a role in this. Our work thus suggest that rotational dynamics are a potentially unifying characteristic through which the often suggested similarity among amorphous materials can be understood.\\ 

\begin{acknowledgments}
We thank the organizers and participants of the Lorentz Center workshop ``Granular Matter Across Scales'' for fostering an environment where the seeds for this work were planted. We are grateful to the late Robert Behringer for always reminding us of the importance of grain rotations and friction. AM and NN are supported in part by the United States National Science Foundation grant CMMI-1727433 and EEC-1840432 (which also involves SS). YL is supported by the University of Minnesota Doctoral Dissertation Fellowship.
\end{acknowledgments}

\clearpage

\section{Methods}
\subsection{Experimental Setup}
In our experiments, we analyze a series of experiments that allow for the tracking of rotation of every disk-shaped particle in a $\sim$1000-particles large shear environment in which shear bands and other large scale inhomogeneities have been completely eliminated, as reported elsewhere \cite{ren13prl,2018_trimer_dong}. Shear is applied quasi-statically from an isotropic stress free state and tracked during the initial shear transient up to a strain of 0.5. Previous experiments have described dilatancy and displacement dynamics in these packings \cite{ren13prl,2018_trimer_dong}. Within the scope of the experimentation in the current research, we study the effect of inter-particle friction using granular assemblies with controlled variations of friction coefficient, as well as the effect of different initial packing fractions upon the response of the granular assembly. One set of particles was cut from photoelastic sheets as in previous experiments\cite{ren13prl}, having an inter-particle friction coefficient $\mu_m$ of approximately 0.7. After conducting experiments with this set, we wrapped these particles with teflon tape. Dry teflon-teflon contacts have a friction coefficient of $\mu_l \sim$0.15~\cite{2020_wang}. A third set of data was obtained with photoelastic disks cut with fine teeth on their circumference so that particles will interlock when they come into contact. Such a particle shape mimics an extremely large friction coefficient; we refer to these particles as $\mu_h$. The diameter ratio of big to small disks is 1.25:1, and the number ratio is roughly 1:3.3 (big to small) for each packing.  Particles were first randomly placed in the shear cell and manually relaxed until no inter-particle contact force very visible by eye. Then starting from either a parallelogram or a rectangle, the shear cell was deformed by strain steps of 0.0027. The system was then relaxed for 10 seconds followed by taking three kinds of pictures: one with white light, one with polarized light, and one with UV light. These three pictures reveal particle positions, particle contact forces/pressure, and particle orientation, respectively. Such a process of shearing, relaxing and picture taking was repeated until a certain total shear strain was achieved. For each packing fraction and friction coefficient, we repeated the experiment five times with the exception of the lowest density $\mu_l$ runs. Note that the analysis of the images acquired during the experiment reveals that not all the grains were detected in all frames, where some grains move out of or inside the boundaries of the images from one frame to another. As a result, for the analysis performed in the current paper, only grains common between all the frames were considered, and the grains present at one frame and not detected in another frame are excluded. Moreover, the grains on the boundary were removed from the analysis.

\subsection{Rotations}
The rigid body rotation between any two frames can be measured as half of the difference in the slope of straight lines fitted to the coordinates of grains centroids in the two frames. We note that, in general, the relation between the measured change in slope and the rigid rotation is nonlinear especially in finite deformation. However, a linear relation in the current analysis for the considered shear strain range is a good approximation.

\subsection{Neighborhood Variance}
The neighborhood variance of each particle refers to the product of its micro-rotation deviation from mean micro-rotation and the mean micro-rotation deviation of its Voronoi neighborhood from mean micro-rotation. We compute the ``average neighborhood variance'' by 
\begin{equation}
    S_{n} = \frac{Z^TWZ}{N-1},\text{ and }Z = \Theta^m - \langle \theta_i^m \rangle,
\end{equation}
where $\Theta^m$ is a vector of particles' micro-rotation whose $i$th element is $\theta_i^m$, $\langle \theta_i^m \rangle$ is the mean of all particles' micro-rotation, and $N$ is the number of particles. $W$ is the row-wise normalized spatial weight matrix. A commonly used spatial weight matrix is the adjacency matrix whose element at the $i$th row and $j$th column indicates whether the $i$th particle is adjacent to the $j$th particle. If the particles are adjacent, the element is 1. Otherwise, the element is 0. A row-wise normalized spatial weight matrix is gotten by dividing each row of a spatial weight matrix by the row sum of the matrix. We conducted analyses of the materials with different surfaces and density: $\mu_l$; $\phi=0.783,0.810,0.828$; $\mu_m$; $\phi=0.692, 0.758,0.816$; $\mu_h$; $\phi=0.713,0.744,0.807$. The observations of each grain were the micro-rotation. We constructed Delaunay triangles to link grains with their neighbor grains, and removed the link whose length was greater than the sum of the radius of the two grains connected by the link.

\subsection{Global Moran's $I$}
Spatial autocorrelation is a measure of the correlation between spatially proximate observations. Positive spatial autocorrelation is the tendency for spatially proximate observations to be similar, while negative spatial autocorrelation means spatially proximate observations tend to be different. Global Moran's $I$ is defined as follows:
\begin{equation}
    I = \frac{Z^TWZ}{Z^TZ} = \sigma_n/\sigma_m, \text{ and } Z=\Theta^m-\langle \theta_i^m \rangle,
\end{equation}
where $\Theta^m$ is a vector of particles' micro-rotation whose $i$th element is $\theta_i^m$, $\langle \theta_i^m \rangle$ is the mean of all particles' micro-rotation, which is negligible. $W$ is the the row-wise normalized spatial weight matrix. This metric measures the average spatial autocorrelation of the entire dataset.  The expected value of global Moran's $I$ under the null hypothesis of no spatial autocorrelation is $E(I)=-\frac{1}{N-1}$, where $N$ is the number of observations. In other words, the more observations there are, the closer the expectation to 0. Values of $I$ usually range from -1 to +1. Values significantly below $E(I)$ indicate negative spatial autocorrelation and values significantly above $E(I)$ indicate positive spatial autocorrelation.

\subsection{Local Moran's $I$}
There are cases where there is no global trend of spatial autocorrelation, but there are local communities where spatial autocorrelation is strong. Local Moran’s $I$ is used to represents the spatial autocorrelation within the local neighborhood of each observation, which is defined as follows:
\begin{equation}
    I_i = \frac{z_i W_{i:}Z}{Z^TZ/(N-1)}, \text{ and } z_i = \theta_i^m - \langle \theta_i^m \rangle,
\end{equation}
where $\theta_i^m$ is the ith particle's micro-rotation. A positive value of $I_i$ means within the $i$th observation’s neighborhood the observations are similar, while a negative value means the observations are different. In order to analyze whether the local communities in a dataset are homogeneous regarding to spatial autocorrelation, we compute the standard deviation of local Moran’s $I$ defined as $\sigma_I$. The greater this standard deviation, the greater the differences between local communities.

\bibliography{apssamp}

\end{document}